\documentclass{PoS}
\usepackage{wrapfig}

\graphicspath{{Figures/}}

\title{Topology, Random Matrix Theory and the spectrum of the Wilson Dirac operator }

\ShortTitle{Topology, Random Matrix Theory and the Wilson Dirac
  operator spectrum}

\author{\speaker{Albert Deuzeman}, Urs Wenger and Ja\"ir Wuilloud\\Albert Einstein Center for Fundamental Physics\\Institute for
Theoretical Physics\\University
of Bern\\Switzerland\\E-mail: \email{deuzeman@itp.unibe.ch}, \email{jair@itp.unibe.ch},
\email{wenger@itp.unibe.ch}}

\abstract{We study the spectrum of the hermitian Wilson Dirac operator
  in the $\epsilon$-regime of QCD in the quenched approximation and
  compare it to predictions from Wilson Random Matrix Theory. Using
  the distributions of single eigenvalues in the microscopic limit and
  for specific topological charge sectors, we examine the possibility
  of extracting estimates of the low energy constants which
  parametrise the lattice artefacts in Wilson chiral perturbation
  theory. The topological charge of the field configurations is
  obtained from a field theoretical definition as well as from the
  flow of eigenvalues of the hermitian Wilson Dirac operator, and we
  determine the extent to which the two are correlated.  }

\FullConference{XXIX International Symposium on Lattice Field Theory\\ July 10 -- 16, 2011\\ Squaw Valley, Lake Tahoe, California}

\begin{document}

\section{Introduction}

It has been known for some time that the low lying eigenvalues of the
QCD Dirac operator are reproduced in the microscopic limit by the
eigenvalues of large random matrices with the correct anti-hermitian
symmetry structure~\cite{Shuryak:1992pi,Verbaarschot:1993pm}. The
connection between random matrix theory (RMT) and chiral perturbation
theory ($\chi$PT) has been established within the so-called
$\epsilon$-regime of QCD, where pion masses are small enough that
their wavelength becomes of the order of the finite size of the
lattice. If this condition is enforced, finite volume effects are
noticeable even when the volume is taken to infinity. Pion momenta are
suppressed and, to leading order, pion fields are constant over the
whole lattice while the effect of spontaneous chiral symmetry breaking
is captured in a single low energy constant (LEC), the quark
condensate $\Sigma$.  From the Banks-Casher relation it is known that
$\Sigma$ can be defined through the spectral density of the Dirac
operator at the origin, so this provides the connection between the
spectrum of the Dirac operator and the one from RMT in the microscopic
regime.

So far, attempts to obtain physical parameters from numerical
simulations of QCD in the $\epsilon$-regime have been restricted to
the use of lattice Dirac operators with an exact chiral
symmetry~\cite{Bietenholz:2003mi,
  Giusti:2003gf,DeGrand:2006uy}. Recent progress has shown, however,
that a sensible analysis can also be performed using the standard
Wilson Dirac operator, for which the chiral symmetry is explicitly
broken.  This is made possible in the framework of Wilson $\chi$PT
that includes the effects of the lattice discretisation and that has
recently been formulated in the
$\epsilon$-regime~\cite{Bar:2008th,Shindler:2009ri}. The Lagrangian of
this effective low-energy theory,
\[
 \mathcal{L}(U) = \frac{1}{2}m \Sigma\mathrm{Tr}(U + U^\dagger) - a^2 W_6 \mathrm{Tr}(U + U^\dagger)^2 - 
    a^2 W_7 \mathrm{Tr}(U - U^\dagger)^2 -a^2 W_8 \mathrm{Tr}(U^2 +
    {U^\dagger}^2) \, ,
\]
contains on top of the continuum Lagrangian three additional operators
which parametrise the lattice artefacts to leading order in $a$ using
the LECs $W_{6,7,8}$. Chiral RMT, too, can be extended to include the
effects of these additional operators, leading to what is known as
Wilson RMT~\cite{Damgaard:2010cz,Akemann:2010em}. There is a
one-to-one correspondence between the new parameters in both
frameworks, so we will use the Wilson $\chi$PT nomenclature for both.

Simulations of QCD, when taken towards the chiral limit at fixed
lattice spacing, are likely to run into numerical instabilities due to
the onset of $\epsilon$-regime dynamics. To avoid such problems, one
needs a thorough understanding of the complicated way the spectrum
depends on the different scales like the quark masses, the lattice
volume and the lattice spacing. In this context, analytic results
concerning the spectral density of the Wilson Dirac operator obtained
in~\cite{Damgaard:2010cz,Akemann:2010em,Kieburg:2011uf} have already
provided important insights.  These proceedings report on a
feasibility study, comparing the Wilson RMT description to results
from quenched QCD simulations in the $\epsilon$-regime. Our primary
interest is whether the effective description can indeed reproduce the
important features of the Wilson Dirac spectrum that we measure, and
we show that this seems to be the case. A secondary goal is to
determine the extent to which one can extract information on the LECs
from this setup. Eventually, the results could also have an impact on
(Wilson) $\chi$PT analysis of dynamical simulations outside of the
$\epsilon$-regime, since the same LECs occur in the regular regimes of
$\chi$PT.

\section{Definitions of topological charge}

The QCD partition function in the $\epsilon$-regime naturally
separates into contributions from different topological charge
sectors~\cite{Leutwyler:1992yt}. When constructing Wilson $\chi$PT in
the $\epsilon$-regime, the presence of the different sectors can be
pa\-ra\-metrised through a phase factor in the partition function,
\[
 Z = \sum\limits_\nu \int\limits_U \mathrm{d}U \det[U]^\nu
 \exp(V\mathcal{L}(U))\, ,
\]
decomposing it in a manner akin to a Fourier
transform~\cite{Damgaard:2010cz}.  As a consequence of this
construction the study of the low lying eigenvalues in the Dirac
operator spectrum can be done separately for each of the charge
sectors, but this obviously requires good control over the topological
properties of the system. However, at finite lattice spacing the
topological charge can not be defined
\begin{wrapfigure}{l}{0.55\textwidth}
   \includegraphics[width=0.55\textwidth]{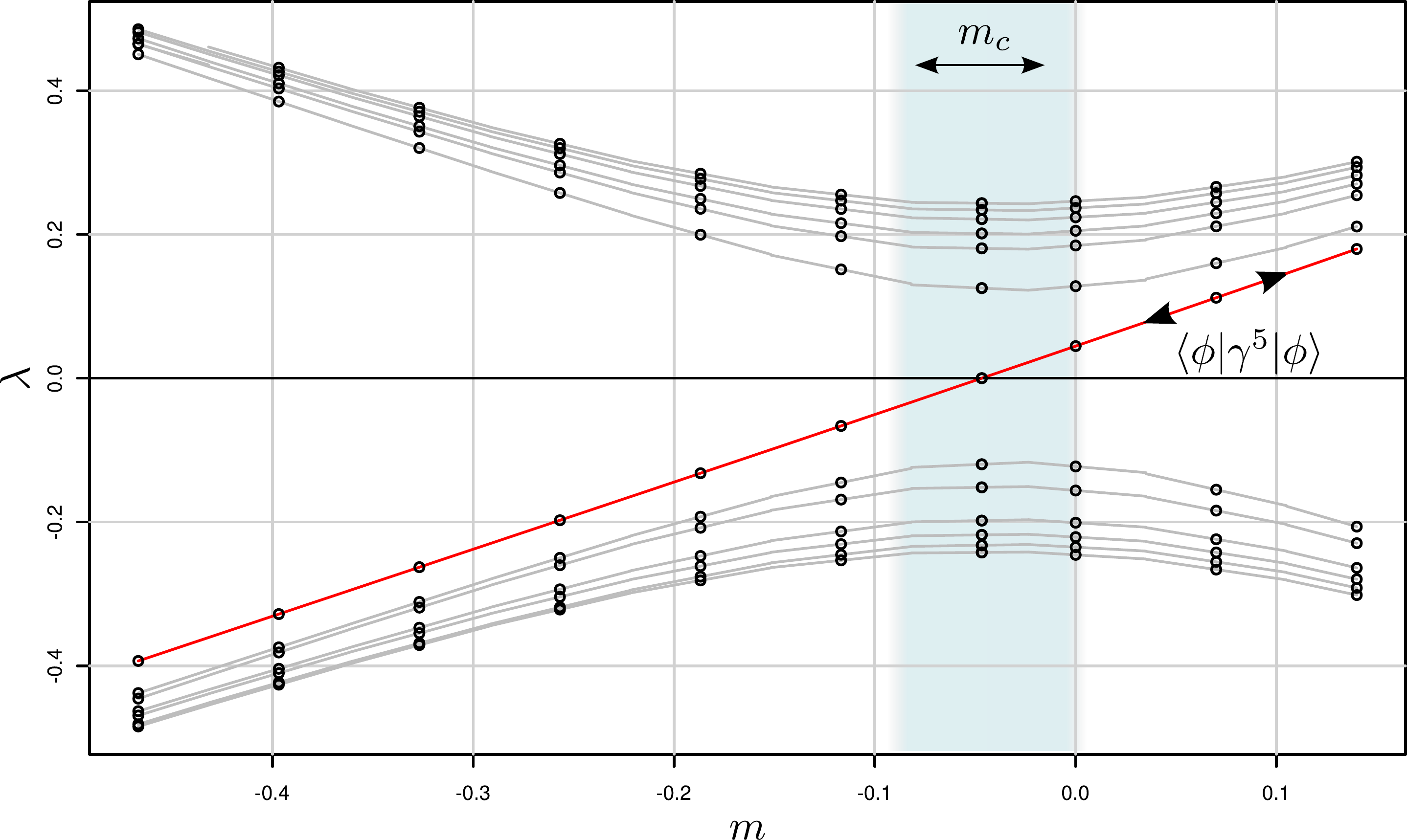}
   \caption{Sample of a typical eigenvalue flow calculation for a volume
   of $24^3\times24$ at a gauge coupling of $\beta=6.2$ using thirty levels of
   HYP smearing. The region around the critical mass $m_c$ where zeros are
   expected to occur is indicated in blue, while the eigenmode drawn in red
   shows a crossing eigenvalue.\label{fig:WF_sample}}
\end{wrapfigure}
uniquely and this creates an ambiguity in the assignment of
configurations to the different charge sectors.  The straightforward
definition from the field strength tensor $\nu_\textrm{\tiny FT} \sim
\int_V F_{\mu\nu} \tilde{F}^{\mu\nu}$ is attractive, because it can
easily be calculated even on large lattices, but for the current
purpose, the natural definition of the charge is through the chiral
overlap index.  One way to calculate it is provided by the eigenvalue
flow method~\cite{Edwards:1998sh, Edwards:1998gk} which counts the
number of real modes of $D_W$ weighted with the signs of their
chiralities. To be specific, the procedure determines those values of
the mass $m$ for which the hermitian operator $D_5(m)$ satisfies
\[
 D_5(m) \psi = \gamma_5 (D_W + m) \psi = 0,
\]
hence the zero modes of $D_5(m)$ correspond to the real eigenvalues of
$D_W$. Moreover, perturbation theory shows that the slope of an
eigenvalue as a function of $m$ is given by the chirality of the
eigenmode, hence the flow of each eigenvalue can be traced as a smooth
function of the mass, as illustrated in figure~\ref{fig:WF_sample},
and the net number of crossings then yields the chiral index. In
practice, one needs to choose a cut-off $m_\mathrm{cut}$ beyond which
no further physically relevant zeros are assumed to exist. Obviously,
this has an influence on the value of the index.  Sufficiently close
to the continuum limit, however, any choice of $m_\mathrm{cut} < m_c
\simeq 0$ will produce the same value for the index. The approach to
this limit can in fact be improved by HYP smearing, and for our
current exploratory studies we rather aggressively use thirty levels
of HYP smearing.

\begin{wrapfigure}{r}{0.400\textwidth}
  \centering
  \includegraphics[width=0.400\textwidth]{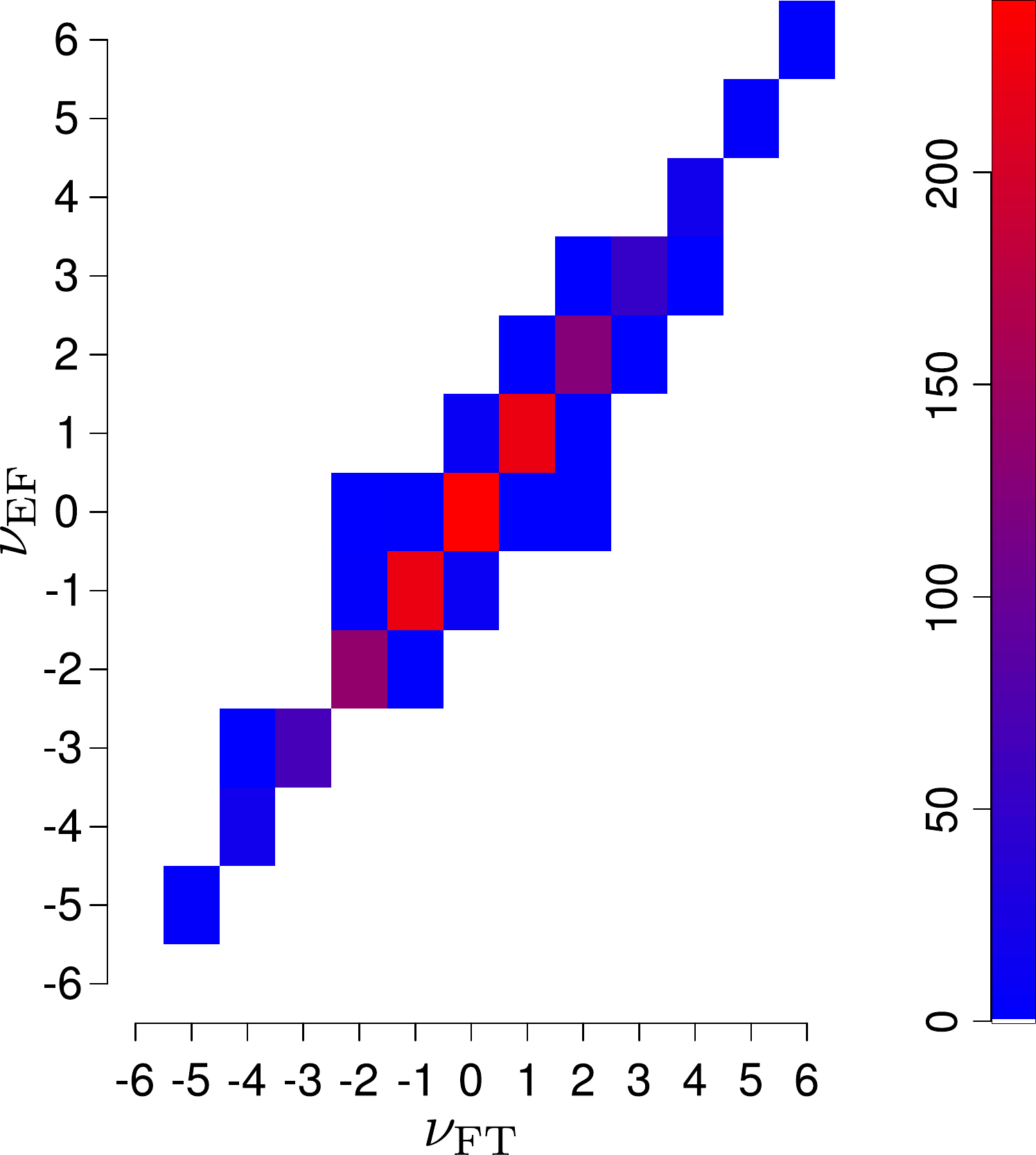}
  \caption{A 2D histogram showing the distribution of the topological
    charge as defined through a field theoretical definition versus
    the charge defined through the eigenvalue flow on a $20^3\times20$
    lattice at $\beta=6.2$.\label{fig:mhist_charge_wf}}
\end{wrapfigure}

\begin{figure}[ht]
  \begin{minipage}[c]{0.500\textwidth}
    \centering
    \begin{tabular}{c|c}
      Parameters & $\mathrm{frac}_{\nu_\mathrm{FT} \neq \nu_\mathrm{EF}}$ \\
      \hline
      $\beta=5.9$, $14^3\times 16$ & 0.065 \\
      $\beta=6.2$, $14^3\times 16$ & 0.008 \\
      $\beta=6.2$, $20^3\times 20$ & 0.036 \\
      $\beta=6.2$, $24^3\times 24$ & 0.096 \\
    \end{tabular}
  \end{minipage}
  \begin{minipage}[c]{0.500\textwidth}
    \centering
    \includegraphics[width=0.9\textwidth]{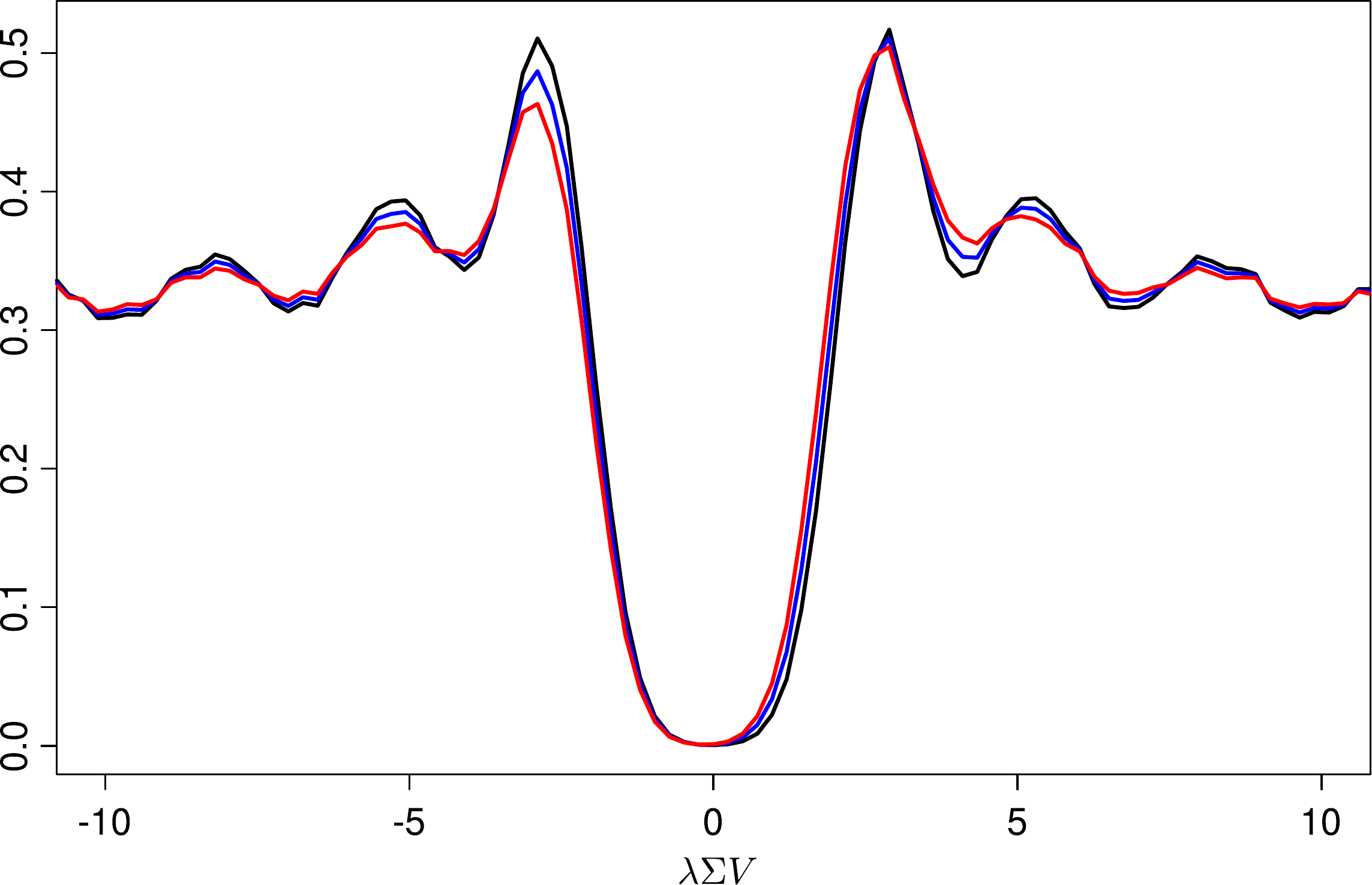}
  \end{minipage}
  \caption{{\it Left:} Table of the estimated fraction of
    configurations that produce a different result for the topological
    charge according to the field theoretical and eigenvalue flow
    definitions. {\it Right:} RMT simulations of the spectral density
    of the Wilson Dirac operator at $\nu=0$, displaying the effect of
    mixing in 0\% (black), 10\% (blue) and 20\% (red) configurations
    with charge $\nu=1$.\label{fig:deviations}}
\end{figure}

A comparison between the field theoretic definition $\nu_\textrm{\tiny
  FT}$ and the one from the eigenflow $\nu_\textrm{\tiny EF}$ is shown
in figure~\ref{fig:mhist_charge_wf} for configurations generated using
the Wilson gauge action at $\beta=6.2$ on a volume of $20^3\times
20$. Both definitions agree for 96.4\% of the configurations, while
the deviations are almost always limited to $\Delta\nu=1$.  The table
included in figure~\ref{fig:deviations} gives estimates for the
fraction of deviating results over a range of parameters. For
relatively fine lattices and moderate volumes, the field theoretical
charge definition and the overlap index agree quite well. The
agreement deteriorates as we go to larger volumes, but the effect is
mild and justifies the use of the field theoretic definition on
lattices for which explicit eigenvalue flow calculations are too
expensive. Of course this comes at the cost of introducing a
systematic error due to the occasional misinterpretation of the
charge.  To estimate the size of that error, one can mimic the effect
of misassigning charges by mixing measurements from two separate RMT
calculations at identical parameters but in different charge sectors.
The results of such an exercise are included in
figure~\ref{fig:deviations}. While the impact of mixing the sectors is
discernible, the impact is small enough to have limited impact on the
precision of our current preliminary fits, even at large mixing ratios
of up to 0.2.

\section{Wilson RMT and the Wilson Dirac operator spectrum of quenched QCD}

With these preliminaries handled, we turn to the fitting of our
lattice data. One prerequisite for a sensible comparison is the
availability of eigenvalues small enough to be within the
$\epsilon$-regime.  When simulating at a Wilson gauge coupling of
$\beta=6.2$, the smallest volume producing a reasonable number of
eigenvalues without apparent bulk effects, but with a mass gap,
turned out to be $24^3\times24$.  At this volume the fraction of
misassigned configurations is a manageable ten percent, so we use the
field teoretical definition of the topological charge instead of the
Wilson flow charge.

\begin{figure}[!ht]
   \begin{center}$
     \begin{array}{cc} 
       \includegraphics[width=0.45\textwidth]{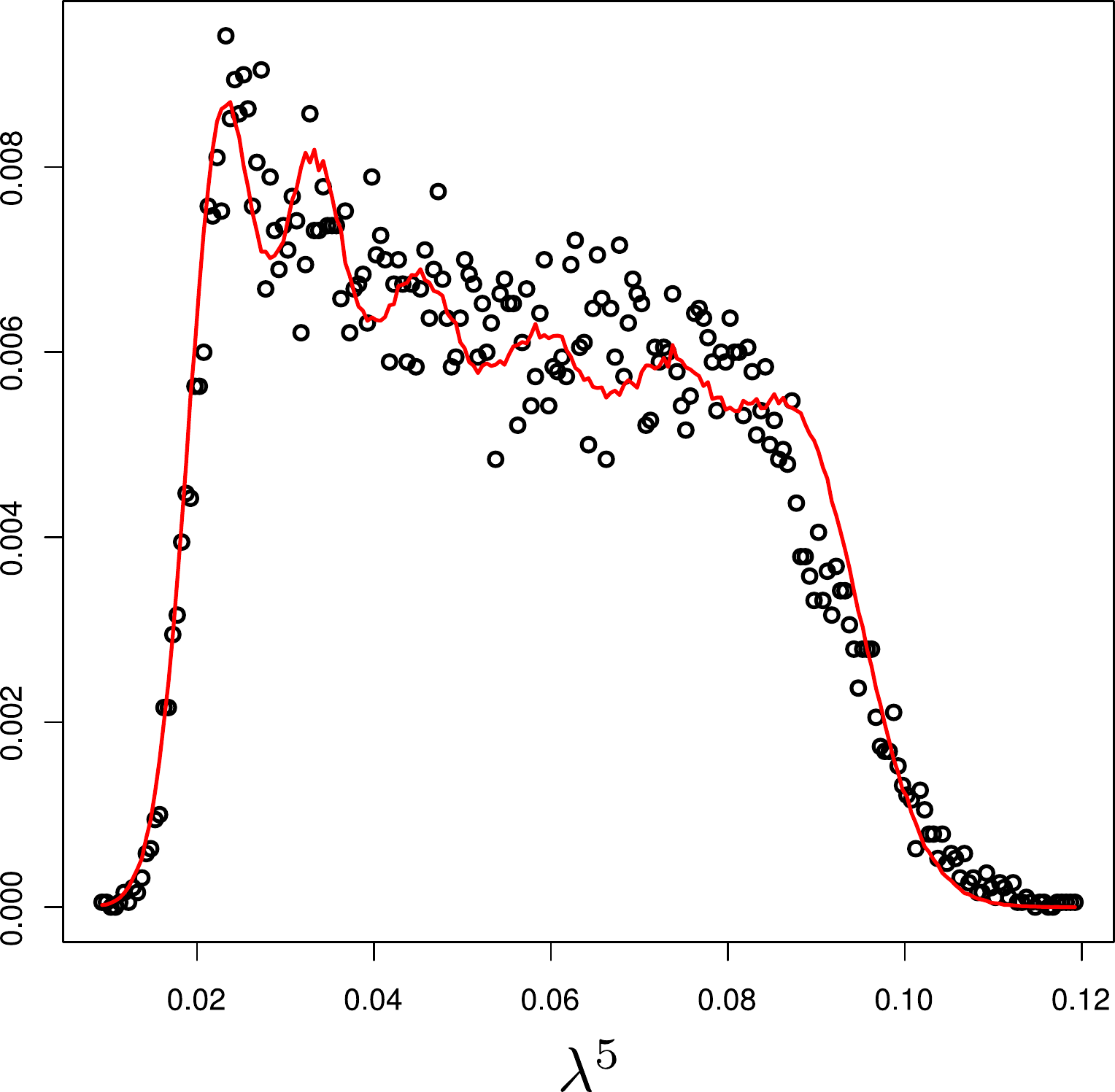} &
       \includegraphics[width=0.45\textwidth]{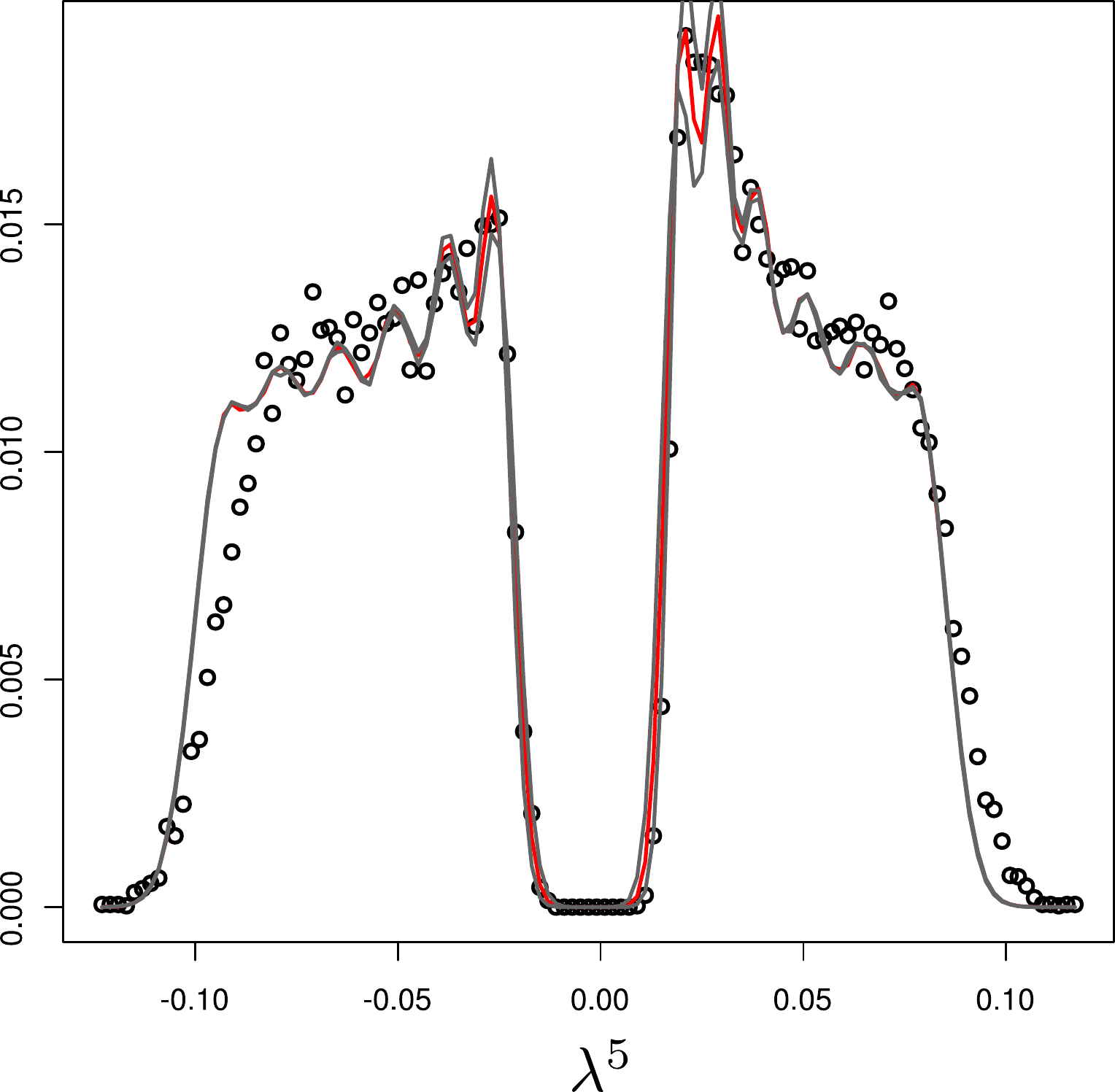}
     \end{array}$
     \caption{Distributions of the 12 lowest lying eigenvalues of the
       hermitian Wilson Dirac
       operator in the sectors $\nu=0$ (left) and $\nu=1$
       (right) together with fits including the effects of the operator
       $W_8$. Results are for a volume of $24^3\times 24$ at
       $\beta=6.2$.  The gray lines in the right panel indicate the
       effects of varying the value of $W_8$ by
       $\pm$10\%.\label{fig:fits}}
   \end{center}
\end{figure}

In figure~\ref{fig:fits} we show two samples of the distribution of
the 12 lowest lying eigenvalues of the Wilson Dirac operator in the
charge sectors $\nu=0$ and 1 for a volume of $24^3\times 24$ at
$\beta=6.2$. The spectrum does not appear to exhibit the undulating
pattern that can, for example, be seen in the RMT data of
figure~\ref{fig:deviations}. If this was down to statistics only, we
would require an increase in statistics by an order of
magnitude. However, rather than using the formulae of
\cite{Akemann:2010em} for the full spectrum, RMT can be used to
extract distributions for the single eigenvalues. This provides
additional information which can be used in the fitting procedure
\cite{Deuzeman:2011dh}.

We therefore implemented a Monte Carlo spectrum calculation using
Wilson RMT.  The spectra were then used to fit the histograms of the
separate eigenvalues, the results of which are also shown in
figure~\ref{fig:fits}. To estimate the precision that can be reached
in the determination of the LECs by such an approach, we vary the
value of $W_8$.  The right panel of figure~\ref{fig:fits} displays
curves showing the effect of a ten percent change in $W_8$ in gray.
The impact of such a change in $W_8$ quickly dissipates beyond the
lowest eigenvalue, however, even a small 10\% variation in $W_8$
produces a clear effect on the spectrum, and we conclude that the
value of $W_8$ can be constrained to within at least ten percent using
this procedure.

\begin{figure}[t]
\centering
   \includegraphics[width=0.45\textwidth]{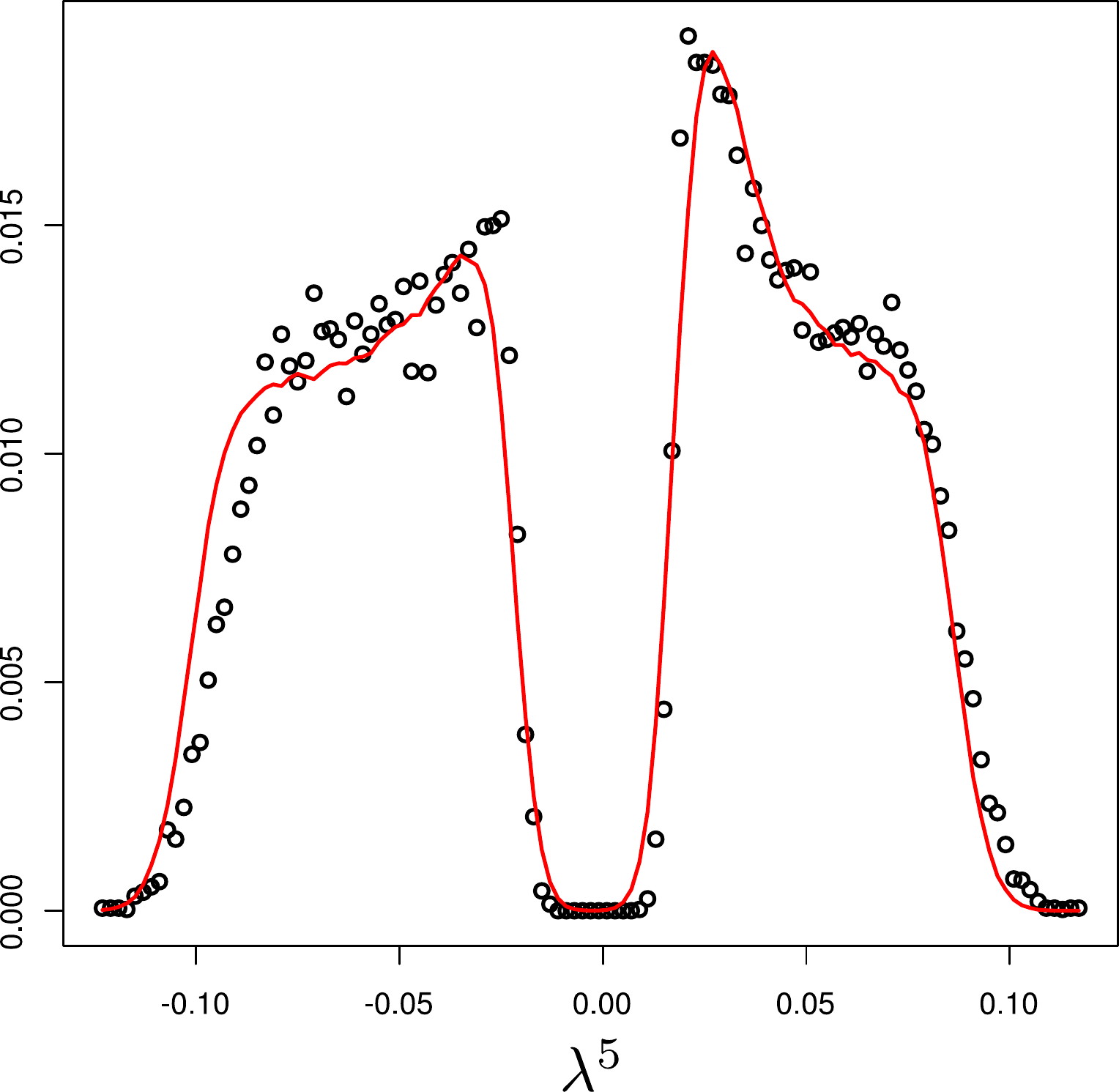}
   \caption{Fits to the low lying eigenvalues of the Wilson Dirac
     operator in the charge sector $\nu=1$, including the effects of
     all three LECs $W_{6,7,8}$.  Results are for a volume of
     $24^3\times 24$ at $\beta=6.2$.\label{fig:global_fit_24_w67}}
\end{figure}

A complicating factor, however, is the potential presence of $W_6$ and
$W_7$ in the effective theory. While the assumption of their
suppression is not unreasonable, they may in fact provide a possible
explanation for the observed lack of structure in the spectra. In our
RMT setup they can straightforwardly be included in the fits and we
refer to~\cite{Akemann:2010em,Deuzeman:2011dh} for details on the
implementation in the Wilson RMT. We find
(cf.~figure~\ref{fig:global_fit_24_w67}) that the optimal fit to
single eigenvalue distributions including now the effects from $W_6,
W_7$ and $W_8$ indeed leads to an absence of structure in the
aggregate spectrum. At this point, however, the effects of the various
LECs
are hard to disentangle and the uncertainties on each of the
parameters is considerably larger than when just taking $W_8$ into
account. Since the separate eigenvalues show a varying sensitivity to
the different LECs, performing such fits at larger volumes, where more
eigenvalues are available within the $\epsilon$-regime, may help
pinning down the various predictions.

\section{Conclusions}

Recent developments which include lattice artefacts into the RMT allow
for detailed predictions of the low lying eigenvalue spectrum of the
Wilson Dirac operator in the $\epsilon$-regime of QCD and provides a
method for determining low energy constants of Wilson chiral
perturbation theory. In these proceedings, we have reported on a
preliminary study concerning the practical applicability of these
predictions and we examined the sensitivity to the LECs.  A
potentially important source of uncertainty for these fits lies in the
correct separation of configurations into topological sectors. We have
studied the use of a field theoretical definition of the topological
charge as a predictor for the overlap index which is the natural
choice in connection with the Wilson Dirac operator spectrum.  A high
degree of correlation between the two definitions is found, although
the field theoretical approach becomes unreliable as the volume is
increased. If the LECs $W_6$ and $W_7$ are assumed to be zero, we find
that $W_8$ can be constrained to within about ten percent. For this
purpose, fits to the separate eigenvalues, currently only available
from direct RMT Monte Carlo calculations, appear to be the most
powerful tool. The particular pattern of deviations seen in the
separate eigenvalues, however, seems to point to a non-negligible
contribution from the additional LECs $W_6$ and $W_7$.  For a more
elaborate and more quantitative analysis along the lines described
here (and consistently using the topological charge as determined from
the eigenvalue flow), we refer to our detailed results presented
in~\cite{Deuzeman:2011dh}. Furthermore, we would also like to point to
the work by Heller et al.~\cite{Damgaard:2011eg} who presented the
results of a similar study at this conference \cite{Heller:2011fg}.

\providecommand{\href}[2]{#2}\begingroup\raggedright\endgroup

\end{document}